\documentclass[twocolumn,showpacs,preprintnumbers,amsmath,amssymb,eqsecnum,showkeys,floatfix,prc]{revtex4}

\usepackage{graphicx}
\usepackage{dcolumn}
\usepackage{bm}

\begin{document}

\newcommand{\vcr}{\mbox{${\bf r}\,$}}
\newcommand{\vcj}{\mbox{${\bf j}\,$}}
\newcommand{\vcJ}{\mbox{${\bf J}\,$}}
\newcommand{\cH}{\mbox{$\cal{H}$}}
\newcommand{\cN}{\mbox{$\cal{N}$}}
\newcommand{\nh}{\mbox{$\hat{n}$}}
\newcommand{\Ph}{\mbox{$\hat{P}$}}
\newcommand{\kh}{\mbox{$\hat{k}$}}
\newcommand{\sh}{\mbox{$\hat{\sigma}$}}
\newcommand{\vchk}{\mbox{$\hat{\bf k}$}}
\newcommand{\nbh}{\mbox{$\hat{\bar{n}}$}}
\newcommand{\pvh}{\mbox{$\hat{\vec{p}}$}}
\newcommand{\magsqr}[1]{\mbox{$\left|{#1}\right|^2$}}
\newcommand{\magabs}[1]{\mbox{$\left|{#1}\right|$}}
\newcommand{\nedots}{\mbox{$.\cdot^{\textstyle{.}}$}}

\preprint{APS/VUHFB-1}

\title{Axially symmetric Hartree-Fock-Bogoliubov Calculations for 
       Nuclei Near the Drip-Lines\\}

\author{E. Ter\'an}
 \email{edgar.teran@vanderbilt.edu}
\author{V.E. Oberacker}%
 \email{volker.e.oberacker@vanderbilt.edu}
\author{A.S. Umar}
 \email{umar@compsci.cas.vanderbilt.edu}
\affiliation{%
Department of Physics and Astronomy, Vanderbilt University, 
Nashville, TN 37235, USA\\
}%

\date{\today}
\begin{abstract}
Nuclei far from stability are studied by solving the Hartree-Fock-Bogoliubov (HFB)
equations, which describe the self-consistent mean field theory with pairing 
interaction. Calculations for even-even nuclei are carried out on a 
two-dimensional axially symmetric lattice, in coordinate space. The quasiparticle 
continuum wavefunctions are considered for energies up to 60 MeV.
Nuclei near the drip lines have a strong coupling between weakly bound 
states and the particle continuum.
This method gives a proper description of the ground state properties of such nuclei.
High accuracy is achieved by representing the operators and wavefunctions using 
the technique of basis-splines. The detailed representation of the HFB equations in cylindrical 
coordinates is discussed.
Calculations of observables for nuclei near the neutron drip line are presented to 
demonstrate the reliability of the method.
\end{abstract}
\maketitle
\section{\label{sec:intro}Introduction\protect\\}

The latest experimental developments as well as recent advances in
computational physics have sparked renewed interest in nuclear
structure theory. In contrast to the well-understood behavior near the
valley of stability, there are many open questions as we move towards
the proton and neutron driplines and towards the limits in mass number
(superheavy region). The neutron dripline represents mostly ``terra
incognita''. In these exotic regions of the nuclear chart, one expects to see several
new phenomena \cite{ISOL97,RIA1}: near the neutron dripline, the
neutron-matter distribution will be very diffuse and of large size
giving rise to ``neutron halos'' and ``neutrons skins''. There are also
expected collective modes associated with this neutron skin, e.g.
the ``scissors'' vibrational mode or the ``pygmy'' resonance. In
proton-rich nuclei, we have recently seen both spherical and deformed
proton emitters; this ``proton radioactivity'' is  caused by the
tunneling of weakly bound protons through the Coulomb barrier.
With RIB facilities, nuclear theorists see an opportunity to study the effective N-N
interaction at large isospin, as well as large pairing correlations.

It is generally acknowledged
that an accurate treatment of the pairing interaction is essential for
describing exotic nuclei \cite{DF84,DN96}.
This work is specifically aimed to calculating
ground state observables such as the total binding energy, charge radii,
proton and neutron densities, separation energies for neutrons and protons,
and pairing gaps.
There are several types of approaches in nuclear structure theory
\cite{RIA1}: for the lightest nuclei, ab-initio calculations
(Green's function Monte Carlo, no-core shell model) based
on the bare N-N interaction are possible \cite{NB97}. Medium-mass nuclei up to
$A \sim 60$ may be treated in the large-scale shell model approach
\cite{KDL97}. For heavier nuclei one utilizes either nonrelativistic
\cite{DF84,DD85,Neg82} or relativistic \cite{Ser92,Rin96,PV97} mean
field theories. 
The large pairing correlations near the driplines can no longer be described
by a small residual interaction. It becomes necessary to treat the
mean field and the pairing field in a single self-consistent
theory.
Furthermore, the outermost nucleons are weakly bound, which implies a
large spatial extent, and they are strongly coupled to the particle
continuum. These features represent major challenges for the mean
field theories. We overcome these difficulties by solving the HFB equations
for deformed, axially symmetric even-even nuclei on a two-dimensional lattice,
without any further approximations. 
So far, most of HFB calculations are based on spherical symmetry or up to a
limited energy in the quasiparticle spectrum continuum.
The importance of the axially symmetry lies on the ability to emulate
a big range of nuclei that are not spherical in nature, e.g. nuclei that
have an non-trivial intrinsic deformation.
We have developed and tested a new mean-field nuclear structure code that
specifically addresses the computational challenges and opportunities 
presented by nuclei near the driplines.

The present work represents an introduction of the splines method to the
solution of the HFB approach in axial symmetry. For now,
we will focus on the methodology of our approach.
We outline here briefly the theoretical and computational details. We also
present results for a few nuclear systems to demonstrate the convergence 
of the results. 

\section{\label{sec:HFB_formalism}Standard HFB Formalism\protect\\}

The many-body Hamiltonian in occupation number representation has the form
\begin{eqnarray}
\hat{H} &=& \sum_{i,j} < i|\ t\ |j> \ \hat{c}_i^\dagger \ \hat{c}_j \nonumber\\
&+& \frac{1}{4} \sum_{i,j,m,n} <ij|\ \bar{v}^{(2)} \ |mn>
\ \hat{c}_i^\dagger \ \hat{c}_j^\dagger \ \hat{c}_n \ \hat{c}_m \ .
\end{eqnarray}
The general linear transformation from particle operators $\hat{c},\hat{c}^\dagger$ 
to quasiparticle operators $\hat{\beta},\hat{\beta}^\dagger$ take the form \cite{RS80}:
\begin{equation}
\left( 
\begin{array}{c}
\hat{\beta} \\ 
\hat{\beta}^\dagger 
\end{array}
\right) =
\left( 
\begin{array}{cc}
 U^\dagger & V^\dagger \\ 
  V^T & U^T 
\end{array}
\right) 
\left( 
\begin{array}{c}
\hat{c} \\ 
\hat{c}^\dagger
\end{array}
\right) \ .
\end{equation}
\noindent The HFB approximate ground state of the many-body system is defined as a
vacuum with respect to quasiparticles
\begin{equation*}
\hat{\beta}_k \ | \Phi_0 > \ = \ 0 \ .
\end{equation*}
\noindent The basic building blocks of the theory are the density matrix
\begin{equation}
\rho_{ij} = < \Phi_0 | \hat{c}_j^\dagger \ \hat{c}_i | \Phi_0 > = (V^*V^T)_{ij}
\ ,
\end{equation}
\noindent and the pairing tensor
\begin{equation}
\kappa_{ij} = < \Phi_0 | \hat{c}_j \ \hat{c}_i | \Phi_0 > = (V^*U^T)_{ij} \ .
\end{equation}
\noindent which give form to the generalized density matrix
\begin{equation*}
\mathcal{R} =
\left( 
\begin{array}{cc}
 \rho & \kappa \\ 
 - \kappa^* & 1 - \rho^* 
\end{array}
\right) \ .
\end{equation*}
\noindent The HFB ground state energy including the constraint on the 
particle number $N$ is given by
\begin{equation}
E({\mathcal{R}}) = < \Phi_0 | \hat{H} - \lambda \hat{N} | \Phi_0 > \ .
\end{equation}
\noindent The equations of motion are derived from 
the variational principle
\begin{equation}
\delta \ [ E( {\mathcal{R}} ) - {\rm{tr}}\  \Lambda ( {\mathcal{R}}^2 - {\mathcal{R}} ) ]  =  0 \ ,
\end{equation}
\noindent which results in the standard HFB formulation
\begin{equation}
[ {\mathcal{H}}, {\mathcal{R}} ] \ = \ 0 \ ,
\end{equation}
\noindent with the generalized single-particle Hamiltonian
\begin{equation}
{\mathcal{H}} =
\left( 
\begin{array}{cc}
 (h - \lambda) & \Delta \\ 
 -\Delta^* & -(h - \lambda)^* 
\end{array}
\right) \ ,
\label{eq:hamilt_matrix}
\end{equation}
where $h$ and $\Delta$ denote the mean field Hamiltonian and pairing potential, 
respectively, and the Lagrange multiplier $\lambda$ is the Fermi energy of the system.

\subsection{Quasiparticle wavefunctions in coordinate space}

In practice, it is to convenient to transform the standard HFB equations into a coordinate 
space representation and solve the resulting differential equations on a lattice. For this 
purpose, we define two types of quasiparticle wavefunctions $\phi_1$ and $\phi_2 $, 
corresponding to each quasiparticle energy state $E_\alpha$:
\begin{subequations}
\begin{eqnarray}
\phi_1^* (E_\alpha, {\bf r} \sigma q) \ &=& \ \sum_i U_{i\alpha} \ (2 \sigma) \ \phi_i ({\bf r} -\sigma q) ,\\
\phi_2 (E_\alpha, {\bf r} \sigma q) \ &=& \ \sum_i V_{i\alpha}^* \ \phi_i ({\bf r} \sigma q) \ .
\end{eqnarray}
\end{subequations}
The basis wavefunctions $\phi_i$ depend on the
coordinate vector ${\bf r}$, the spin projection $\sigma = \pm \frac{1}{2}$, and 
the isospin projection $q$ ($q=+\frac{1}{2}$ corresponds to protons and $q=-\frac{1}{2}$ to
neutrons). \\
The particle density matrix for the HFB ground state assumes a very
simple mathematical structure in terms of $\phi_1$ and $\phi_2 $ \cite{DN96} :
\begin{eqnarray}
\rho ({\bf r} \sigma q, {\bf r}' \sigma' q') \ &=& \ 
< \Phi_0 | \ \hat{\psi}^\dagger ({\bf r}' \sigma' q') \ \hat{\psi} ({\bf r} \sigma q) \ | \Phi_0 > \nonumber \\
&=& \sum_{i,j} \rho_{ij} \ \phi_i ({\bf r} \sigma q) \ \phi_j^* ({\bf r}' \sigma' q') \nonumber\\
&=& \sum_{E_\alpha > 0}^{\infty} 
\phi_2 (E_\alpha, {\bf r} \sigma q) \ \phi_2^* (E_\alpha, {\bf r}' \sigma' q') \ . 
\nonumber\\
\label{eq:particle_density_matrix}
\end{eqnarray}
The sum over the states $E_\alpha$ has replaced the integral form of the equations, 
since the HFB continuous spectrum has been discretized for practical calculations.\\
Instead of the standard antisymmetric pairing tensor $\kappa$ defined as
\begin{equation}
\kappa ({\bf r} \sigma q, {\bf r}' \sigma' q') \ = \ 
< \Phi_0 | \ \hat{\psi} ({\bf r}' \sigma' q') \ \hat{\psi} ({\bf r} \sigma q) \ | \Phi_0 > 
\label{eq:pairing_tensor}
\end{equation}
we introduce the pairing density matrix $\tilde{\rho}$ which is Hermitian for a time-reversal
invariant ground state and hence more convenient to use \cite{DN96} :
\begin{eqnarray}
\tilde{\rho}({\bf r} \sigma q, {\bf r}' \sigma' q') \ &=& \ 
(-2 \sigma') \ \kappa ({\bf r} \sigma q, {\bf r}' -\sigma' q') \nonumber \\
&=& (-2 \sigma') \sum_{i,j} \kappa_{ij} \ \phi_i ({\bf r} \sigma q) \ \phi_j ({\bf r}' -\sigma' q') \nonumber\\
&=& - \sum_{E_\alpha > 0}^{\infty} \phi_2 (E_\alpha, {\bf r} \sigma q) \ 
    \phi_1^* (E_\alpha, {\bf r}' \sigma' q') \ .
\nonumber\\
\label{eq:pairing_density_matrix}
\end{eqnarray}
In principle, the sums go over all the energy states, but in practice a cutoff 
in the number of states is done up to a reasonable number ($\sim
60$ MeV).

Proceeding in analogy to the pairing density matrix, we replace the antisymmetric
pairing potential $\Delta$ in Eq. (\ref{eq:hamilt_matrix}) with the Hermitian
pairing field $\tilde{h}$
\begin{equation}
\tilde{h}({\bf r} \sigma q, {\bf r}' \sigma' q') \ = \ 
(-2 \sigma') \ \Delta ({\bf r} \sigma q, {\bf r}' -\sigma' q') \ .
\end{equation}

\subsection{Normal density and pairing density}

From expressions (\ref{eq:particle_density_matrix}) and (\ref{eq:pairing_density_matrix}) for 
the density  matrices we obtain the following expressions for the normal density 
$\rho_q({\bf r})$ and pairing density $\tilde \rho_q({\bf r})$, which are defined as 
the spin-averaged diagonal elements of their correspondent matrices
\begin{eqnarray}
\rho_q({\bf r}) \ &=& \ \sum_{\sigma} \rho ({\bf r} \sigma q, {\bf r} \sigma q) \nonumber\\
\ &=& \ \sum_{\sigma}
\sum_{\alpha} \phi_{2,\alpha} ({\bf r} \sigma q) \ \phi_{2,\alpha}^* ({\bf r} \sigma q) \ ,
\label{eq:density}    \\
\tilde{\rho_q}({\bf r}) \ &=& \ \sum_{\sigma} \tilde{\rho} ({\bf r} \sigma q, {\bf r} \sigma q) \nonumber\\
\ &=& \ - \sum_{\sigma}
\sum_{\alpha} \phi_{2,\alpha} ({\bf r} \sigma q) \ \phi_{1,\alpha}^* ({\bf r} \sigma q) \ . 
\label{eq:pairing_density} 
\end{eqnarray}
The quasiparticle energy $E_\alpha$ is denoted by index $\alpha$ for simplicity.
The physical interpretation of $\tilde{\rho_q}$ has been discussed in \cite{DN96}:
the quantity $[\tilde{\rho_q}({\bf r})\ \Delta V /2]^2$ gives the probability to find a
\emph{correlated} pair of nucleons with opposite spin projection in the volume
element $\Delta V$. 

\subsection{Kinetic and spin-orbit densities}

The kinetic energy density $\tau_q({\bf r})$ is defined as a functional of 
wavefunctions $\phi_2$
\begin{eqnarray}
\tau_q({\bf r}) \ &=& \ \nabla \cdot \nabla' \rho_q({\bf r},{\bf r}') |_{{\bf r}={\bf r}'} \nonumber\\
\ &=& \ 
\nabla \cdot \nabla' \left( \sum_{\sigma} \rho ({\bf r} \sigma q, {\bf r}' \sigma q) 
\right) |_{{\bf r}={\bf r}'} \nonumber \\
\ &=& \ \sum_{\sigma} \sum_{\alpha} | \nabla \ \phi_{2,\alpha} ({\bf r} \sigma q) |^2 \ .
\label{eq:kinetic_energy}
\end{eqnarray}
The spin-orbit density does not appear directly in the nuclear potential,
but rather its divergence
\begin{equation}
{\bf \nabla \cdot J}_q( {\bf r} ) = -i \sum_{\alpha} 
(\nabla \phi_{2,\alpha}^{\dagger}( {\bf r},q )) \cdot 
(\nabla \times \sigma) \phi_{2,\alpha}( {\bf r},q )\;\;.
\label{eq:current_divergence}
\end{equation}

\subsection{Energy functional and mean fields}

Standard HFB theory yields the following expression for the total binding energy
of the nucleus in its ground state, with contributions from the mean field
and the pairing field
\begin{equation*}
E_{HFB} = < \Phi_{HFB} | \hat{H} | \Phi_{HFB} > = E_{mf} +E_{pair}\ .
\end{equation*}
\noindent To simplify the notation, we drop the isospin indices $q,q'$ in this
section and in the following section. In coordinate space, the mean field contribution
is given by \cite{DN96}
\begin{eqnarray}
E_{mf} &=& \frac{1}{2} \int d^3 r \int d^3 r' \sum_{\sigma,\sigma'} \left[ \ 
 t({\bf r} \sigma, {\bf r}' \sigma') + h({\bf r} \sigma, {\bf r}' \sigma') \ \right] \nonumber\\
&\times& \rho ({\bf r}' \sigma', {\bf r} \sigma) \ ,
\end{eqnarray}
and pairing energy contribution has the form
\begin{equation}
E_{pair} = \frac{1}{2} \int d^3 r \int d^3 r' \sum_{\sigma,\sigma'}
  \tilde{h}({\bf r} \sigma, {\bf r}' \sigma') \  \tilde{\rho} ({\bf r}' \sigma',
   {\bf r} \sigma) \ .
\label{eq:pairing_energy}
\end{equation}
The quantity $h$ denotes the mean field, i.e. the particle-hole (p-h) channel of the interaction
\begin{widetext}
\begin{equation}
h({\bf r} \sigma, {\bf r}' \sigma') \ = \ t({\bf r} \sigma, {\bf r}' \sigma') \ + \ 
\int d^3 r_2 \int d^3 r_2' \sum_{\sigma_2,\sigma_2'} \ \bar{v}^{(2)}({\bf r} \sigma, {\bf r_2}
\sigma_2; {\bf r}' \sigma', {\bf r_2}' \sigma_2') \  
\rho ({\bf r_2}' \sigma_2', {\bf r_2} \sigma_2) \ .
\end{equation}
\end{widetext}
where ${\bar{v}_{12}}^{(2)}$ is the antisymmetrized two-body effective N-N interaction (see Appendix).
\noindent The kinetic energy matrix elements are given by
\begin{equation}
t({\bf r} \sigma, {\bf r}' \sigma') \ = \ 
\delta({\bf r} - {\bf r}') \ \delta_{\sigma,
\sigma'} \ \left(- \frac{\hbar^2}{2m} \ \nabla^2 \right) \ .
\label{eq:kinetic_energy_matrix}
\end{equation}
In a similar way, we find for the pairing mean field $\tilde{h}$, i.e. for the p-p and h-h
channels of the interaction
\begin{widetext}
\begin{equation}
\tilde{h}({\bf r} \sigma, {\bf r}' \sigma') \ = \ 
\int d^3 r_1' \int d^3 r_2' \sum_{\sigma_1',\sigma_2'} \ 2 \sigma' \sigma_2' \ 
\bar{v}^{(2)}_{pair} ({\bf r} \sigma, {\bf r}' - \sigma'; {\bf r_1}' \sigma_1', {\bf r_2}' - \sigma_2') \  
\tilde{\rho} ({\bf r_1}' \sigma_1', {\bf r_2}' \sigma_2') \ .
\end{equation}
\end{widetext}


\subsection{Pairing interaction}

In practice, one tends to use \emph{different} effective N-N interactions for
the p-h and for the p-p channel. If one assumes that the effective interaction $\bar{v}^{(2)}_{pair}$ is local
\begin{eqnarray*}
\bar{v}^{(2)}_{pair} ({\bf r} \sigma, {\bf r}' - \sigma'; {\bf r_1}' \sigma_1', {\bf r_2}' - \sigma_2') \ =
\ \delta({\bf r_1}' - {\bf r}) \ \delta_{\sigma_1',\sigma} \\
\times \ \delta({\bf r_2}' - {\bf r}') \ \delta_{\sigma_2',\sigma'}
V_p({\bf r} \sigma, {\bf r}' - \sigma') \ ,
\end{eqnarray*}
the pairing mean field Hamiltonian becomes
\begin{equation*}
\tilde{h}({\bf r} \sigma, {\bf r}' \sigma') \ = \ V_p({\bf r} \sigma, {\bf r}' - \sigma') \ 
\tilde{\rho} ({\bf r} \sigma, {\bf r}' \sigma') \ .
\end{equation*}
For the pairing interaction $V_p$ we utilize the form
\begin{equation*}
V_p({\bf r} \sigma, {\bf r}' - \sigma') \ = \ V_0 \ \delta({\bf r} - {\bf r}')
            \ \delta_{\sigma,\sigma'} \ F({\bf r}) \ .
\end{equation*}
This parameterization describes two primary pairing forces: 
a pure delta interaction ($F=1$) that gives rise to {\it volume pairing}, and a 
density dependent delta interaction (DDDI) that gives rise to {\it surface pairing}. 
In the latter case, one uses the following phenomenological ansatz \cite{RD99}
for the factor $F$
\begin{equation}
F({\bf r}) \ = \ 1 - \left( \frac{\rho({\bf r})}{\rho_0}  \right)^\gamma
\end{equation}
where $\rho({\bf r})$ is the mass density. 

The DDDI interaction generates the following pairing mean field for the two
isospin orientations $q = \pm \frac{1}{2}$
\begin{equation}
\tilde{h}_q({\bf r} \sigma, {\bf r}' \sigma') \ = \ \frac{1}{2} \ V_0^{(q)} 
         \tilde{\rho_q}({\bf r}) F({\bf r})
         \ \delta({\bf r} - {\bf r}') \ \delta_{\sigma,\sigma'} \;\; .
\end{equation}
The pairing contribution to the nuclear binding energy is then
\begin{eqnarray*}
E_{pair} &=& E_{pair}^{(p)} + E_{pair}^{(n)} \\
&=& \int d^3 r
    \left[ \frac{V_0^{(p)}}{4} \tilde{\rho}_p^{\ 2} ({\bf r}) 
  + \frac{V_0^{(n)}}{4} \tilde{\rho}_n^{\ 2} ({\bf r}) \right] F({\bf r}) \ .
\end{eqnarray*}
An important related quantity is the average pairing gap for protons and
neutrons which is defined as \cite{DF84,DN96}
\begin{eqnarray*}
<\Delta_q> \ = \ - \frac{1}{N_q} \ trace \left( \tilde{h}_q \ \rho_q \right)
\nonumber \\
= \ - \frac{1}{N_q} \ \int d^3 r \int d^3 r' \sum_{\sigma,\sigma'}
  \tilde{h}_q ({\bf r} \sigma, {\bf r}' \sigma') \  \rho_q ({\bf r}' \sigma',
   {\bf r} \sigma)
\end{eqnarray*}
where $N_q$ denotes the number of protons or neutrons. Inserting the expression
derived earlier for the mean pairing field we arrive at
\begin{equation}
<\Delta_q> \ = \ - \frac{1}{2} \ \frac{V_0^{(q)}}{N_q} \int d^3 r \ 
\tilde{\rho_q}({\bf r}) \ \rho_q({\bf r}) \ F({\bf r}) \ .
\label{eq:mean_delta}
\end{equation}
Note that the pairing gap is a positive quantity because $V_0^{(q)}<0$.


\subsection{\label{sec:hfb_eqns}HFB equations in coordinate space}

For certain types of effective interactions (e.g. Skyrme mean field and pairing
delta-interactions) the particle Hamiltonian $h$ and the
pairing Hamiltonian $\tilde h$ are diagonal in isospin space and
local in position space,
\begin{subequations}
\begin{equation}
h({\bf r} \sigma q, {\bf r}' \sigma' q') \ = \ \delta_{q,q'}
      \ \delta({\bf r} - {\bf r}') h^q_{\sigma, \sigma'}({\bf r}) 
\end{equation}
and
\begin{equation}
\tilde{h}({\bf r} \sigma q, {\bf r}' \sigma' q') \ = \ \delta_{q,q'}
      \ \delta({\bf r} - {\bf r}') \tilde{h}^q_{\sigma, \sigma'}({\bf r}) \ .
\end{equation}
\end{subequations}
Inserting these into the above HFB equations results in
a 4x4 structure in spin space:
\begin{equation}
\left( 
\begin{array}{cc}
( h^q -\lambda ) & \tilde h^q \\
\tilde h^q & - ( h^q -\lambda ) 
\end{array}
\right)
\left(
\begin{array}{c}
\phi^q_{1,\alpha} \\  
\phi^q_{2,\alpha}
\end{array}
\right)
 = E_\alpha
\left( 
\begin{array}{c}
\phi^q_{1,\alpha} \\  
\phi^q_{2,\alpha}
\end{array}
\right)
\label{eq:hfbeq2}
\end{equation}
with
\begin{equation*}
h^q =
\left( 
\begin{array}{cc}
h^q_{\uparrow \uparrow}({\bf r}) & h^q_{\uparrow \downarrow}({\bf r}) \\
h^q_{\downarrow \uparrow}({\bf r}) & h^q_{\downarrow \downarrow}({\bf r})
\end{array}
\right)
, \  
\tilde h^q =
\left( 
\begin{array}{cc}
\tilde h^q_{\uparrow \uparrow}({\bf r}) & \tilde h^q_{\uparrow \downarrow}({\bf r}) \\ 
\tilde h^q_{\downarrow \uparrow}({\bf r}) & \tilde h^q_{\downarrow \downarrow}({\bf r})
\end{array}
\right) \ \ .
\label{hspin}
\end{equation*}
Because of the
structural similarity between the Dirac equation and the HFB equation in
coordinate space, we encounter here similar computational challenges: for
example, the spectrum of quasiparticle energies $E$ is unbounded from above {\em
and} below. The spectrum is discrete for $|E|<-\lambda$
and continuous for $|E|>-\lambda$. For even-even nuclei it is customary to 
solve the HFB equations with a positive
quasiparticle energy spectrum $+E_\alpha$ and consider all negative
energy states as occupied in the HFB ground state.


\section{\label{sec:axial_eqns}2-D Reduction for Axially Symmetric Systems}

For simplicity, we assume that the HFB quasiparticle Hamiltonian
is invariant under rotations ${\hat R}_z$ around
the z-axis, i.e. $[{\mathcal{H}},{\hat R}_z]=0$. Due to the axial symmetry
of the problem, it is advantageous to introduce cylindrical coordinates
$(\phi,r,z)$.
It is possible to construct simultaneous
eigenfunctions of the generalized Hamiltonian ${\mathcal{H}}$ and
the z-component of the angular momentum, ${\hat j}_z$
\begin{subequations}
\begin{eqnarray}
{\mathcal{H}} \ \psi_{n,\Omega,q} (\phi,r,z) & = & E_{n,\Omega,q} \ 
    \psi_{n,\Omega,q} (\phi,r,z)  \\
{\hat j}_z \ \psi_{n,\Omega,q} (\phi,r,z) & = & \hbar \Omega \ 
    \psi_{n,\Omega,q} (\phi,r,z) \ ,
\end{eqnarray}
\label{eq:angular_momentum}
\end{subequations}
with quantum numbers 
$\Omega = \pm \frac{1}{2}, \pm \frac{3}{2},\pm \frac{5}{2}, ...$ corresponding to
each $nth$ energy state.
The simultaneous quasiparticle eigenfunctions take the form
\begin{eqnarray}
\psi_{n,\Omega,q} (\phi,r,z) &=& 
\left( 
\begin{array}{c}
\phi^{(1)}_{n,\Omega,q} (\phi,r,z) \\ 
\phi^{(2)}_{n,\Omega,q} (\phi,r,z) \\ 
\end{array} 
\right) \nonumber\\ 
&=&
\frac{1}{\sqrt{2 \pi}}
\left( 
\begin{array}{c}
e^{i(\Omega - \frac 12)\phi} \ \phi^{(1)}_{n,\Omega,q} (r,z,\uparrow) \\ 
e^{i(\Omega + \frac 12)\phi} \ \phi^{(1)}_{n,\Omega,q} (r,z,\downarrow) \\
e^{i(\Omega - \frac 12)\phi} \ \phi^{(2)}_{n,\Omega,q} (r,z,\uparrow) \\ 
e^{i(\Omega + \frac 12)\phi} \ \phi^{(2)}_{n,\Omega,q} (r,z,\downarrow)
\end{array} 
\right) \ .
\label{eq:wvfnctn}
\end{eqnarray}
We introduce the following useful notation
\begin{subequations}
\begin{eqnarray}
U^{(1,2)}_{ n \Omega q} (r,z) =  \phi^{(1,2)}_{n,\Omega,q} (r,z,\uparrow)\ , \\ 
L^{(1,2)}_{ n \Omega q} (r,z) =  \phi^{(1,2)}_{n,\Omega,q} (r,z,\downarrow) \ .
\end{eqnarray}
\end{subequations}
From the vanishing commutator, $ [ {\mathcal H}, j_z ]$, we
can determine the $\phi$-dependence of the HFB quasiparticle Hamiltonian
and arrive at the following structure for the Hamiltonian
\begin{equation}
h (\phi,r,z) =
\left( 
\begin{array}{cc}
h^{\prime}_{\uparrow \uparrow} \ (r,z) & e^{-i \phi} \ h^{\prime}_{\uparrow \downarrow} \ (r,z) \\ 
e^{+i \phi} \ h^{\prime}_{\downarrow \uparrow} \ (r,z) & h^{\prime}_{\downarrow \downarrow} \ (r,z)
\end{array}
\right) \ .
\label{eq:hamil}
\end{equation}
and the pairing Hamiltonian
\begin{equation}
\tilde h (\phi,r,z) =
\left( 
\begin{array}{cc}
\tilde h^{\prime}_{\uparrow \uparrow} \ (r,z) & e^{-i \phi} \ \tilde h^{\prime}_{\uparrow \downarrow} \ (r,z) \\ 
e^{+i \phi} \ \tilde h^{\prime}_{\downarrow \uparrow} \ (r,z) & \tilde h^{\prime}_{\downarrow \downarrow} \ (r,z)
\end{array}
\right) \ ,
\label{eq:hamil-tilda}
\end{equation}
Inserting equations (\ref{eq:hamil}) and  (\ref{eq:hamil-tilda}) into the eigenvalue 
Eq. (\ref{eq:hfbeq2}), we arrive at the \emph{reduced 2-D problem}
in cylindrical coordinates:
\begin{eqnarray}
\left( \begin{array}{cccc}
(h'_{\uparrow \uparrow} - \lambda) & h'_{\uparrow \downarrow} &
 \tilde{h'}_{\uparrow \uparrow} & \tilde{h'}_{\uparrow \downarrow} \\
 h'_{\downarrow \uparrow} & (h'_{\downarrow \downarrow} - \lambda) &
 \tilde{h'}_{\downarrow \uparrow} & \tilde{h'}_{\downarrow \downarrow} \\
 \tilde{h'}_{\uparrow \uparrow} & \tilde{h'}_{\uparrow \downarrow} &
 -(h'_{\uparrow \uparrow} - \lambda) &  -h'_{\uparrow \downarrow} \\
 \tilde{h'}_{\downarrow \uparrow} & \tilde{h'}_{\downarrow \downarrow} &
 -h'_{\downarrow \uparrow} & -(h'_{\downarrow \downarrow} - \lambda) 
\end{array} \right) \nonumber\\
\times
\left( \begin{array}{c}
                U^{(1)}_{n,\Omega,q} \\
                L^{(1)}_{n,\Omega,q} \\
                U^{(2)}_{n,\Omega,q} \\
                L^{(2)}_{n,\Omega,q} 
\end{array} \right) 
 = E_{n,\Omega,q} \ 
\left( \begin{array}{c}
                U^{(1)}_{n,\Omega,q} \\
                L^{(1)}_{n,\Omega,q} \\
                U^{(2)}_{n,\Omega,q} \\
                L^{(2)}_{n,\Omega,q}  
\end{array} \right)
\label{eq:hfb2d}
\end{eqnarray}
Here, quantities $\tilde{h'}$, $h'$, $U$ and $L$ are all functions of $(r,z)$ only.
Also, $\tilde{h'}$ and $h'$ contain the implicit isospin dependence $q$. 
This is the main mathematical structure that we implement in computational calculations.
For a given angular momentum projection quantum number $\Omega$, we solve
the eigenvalue problem to obtain energy 
eigenvalues $E_{n,\Omega,q}$ and eigenvectors $\psi_{n,\Omega,q}$
for the corresponding HFB quasiparticle states. 

\subsection{Representation of operators}

The Hartree--Fock Hamiltonian using the Skyrme effective interaction can be written
\begin{eqnarray}
h_q = &-& {\bf\nabla  \cdot} \frac{\hbar^2}{2 m_q^*} {\bf \nabla} 
    + U_q + U_C\delta_{q\frac{1}{2}} \nonumber\\
    &+& \frac{1}{2 i} \left( {\bf \nabla \cdot I}_q 
   + {\bf I}_q{\bf \cdot \nabla}
    \right)  - i {\bf B}_q{\bf \cdot}\left( {\bf \nabla \times \sigma} \right)\;\;.
\end{eqnarray}
where $U_q$ is the nuclear central field, $U_C$ the Coulomb interaction,
and the spin-orbit field part is given by 
${\bf B}_q{\bf \cdot}\left( {\bf \nabla \times \sigma} \right)$.
The explicit form of these expressions for the case of
the Skyrme interaction are included in the Appendix.
Starting from the kinetic energy we apply the cylindrical form of the Laplacian operator to the standard
form of the wavefunction in Eq.(\ref{eq:wvfnctn}), to find
\begin{equation}
\hat{t}_q = \left( \begin{array}{cc} t_{11} & 0 \\ 0 & t_{22} \end{array} \right) \ ,
\end{equation}
whose elements are given by
\begin{subequations}
\begin{eqnarray}
t_{11} = 
 &f& \left( \frac{\partial^2}{\partial r^2} 
 + \frac{1}{r} \frac{\partial}{\partial r}
 - \left(\frac{(\Omega - 1/2)}{r}\right)^2 
 + \frac{\partial^2}{\partial z^2} \right) \nonumber\\
 &+& \frac{\partial f}{\partial r} \frac{\partial}{\partial r}
 + \frac{\partial f}{\partial z} \frac{\partial}{\partial z} \\
t_{22} = 
 &f& \left( \frac{\partial^2}{\partial r^2} 
 + \frac{1}{r} \frac{\partial}{\partial r}
 - \left(\frac{(\Omega + 1/2)}{r}\right)^2 
 + \frac{\partial^2}{\partial z^2} \right) \nonumber\\ 
 &+& \frac{\partial f}{\partial r} \frac{\partial}{\partial r}
 + \frac{\partial f}{\partial z} \frac{\partial}{\partial z} \ ,
\end{eqnarray}
\end{subequations}
$f$ being the effective mass given in (\ref{effective_mass_def}).
The local potential terms could also be cast into a matrix form
\begin{equation}
\hat{v}_q = \left( \begin{array}{cc} v_{11} & 0 \\ 0 & v_{22} \end{array} \right) \ ,
\end{equation}
where 
\begin{equation}
v_{11} = v_{22} = U_q + U_C\delta_{q\frac{1}{2}}\;\;.
\end{equation}
Expressions for $U_q$ and $U_C$ are given in the Appendix.
The Hartree-Fock spin-orbit operator
\begin{equation}
\label{hf_ls} 
- i {\bf B}_q{\bf \cdot}\left( {\bf \nabla \times \sigma} \right) 
\longrightarrow   \hat{w}_q\;\;,
\end{equation}
can similarly be written into the form
\begin{equation}
\hat{w}_q  = \left( \begin{array}{cc} w_{11} & w_{12} \\  
                           w_{21} & w_{22} \end{array} \right) \ ,
\label{eq:spin_orbit_operator}
\end{equation}
with \cite{K96}
\begin{subequations}
\begin{eqnarray}
w_{11} &=& {\cal B}_r \frac{\Omega - 1/2}{r} \\
w_{12} &=& \left[- {\cal B}_z\frac{\Omega+1/2}{r}  -
        {\cal B}_z\frac{\partial}{\partial r} + 
        {\cal B}_r \frac{\partial}{\partial z}\right]  \\
w_{21} &=& \left[-{\cal B}_z \frac{\Omega - 1/2}{r} + 
        {\cal B}_z\frac{\partial}{\partial r} 
        -{\cal B}_r \frac{\partial}{\partial z}\right] \\
w_{22} &=&  - {\cal B}_r\frac{\Omega + 1/2}{r}  \ ,             
\end{eqnarray}
\end{subequations}
where ${\cal B}_r, {\cal B}_z$ are defined in the Appendix for the Skyrme force.

\subsection{\label{sec:level2}Densities and currents}

Making use of the definitions for the normal density and pairing density,
Eqs. (\ref{eq:density}) and (\ref{eq:pairing_density}), we apply the bi-spinor structure of the quasiparticle 
wavefunctions to find the corresponding expressions in axial symmetry:
\begin{widetext}
\begin{eqnarray}
  \rho_q(r,z) &=& \frac{1}{2 \pi} 
  \left(2 \sum_{\Omega>0}^{\Omega_{max}} \right) 
  \times \sum_{E_n>0}^{E_{max}} 
  \left[|U^{(2)}_{n \Omega q}(r,z)|^2  + |L^{(2)}_{n \Omega q}(r,z)|^2 \right] \\
  \tilde{\rho}_q(r,z) &=& - \frac{1}{2 \pi} 
  \left(2 \sum_{\Omega>0}^{\Omega_{max}} \right)  
  \times \sum_{E_n>0}^{E_{max}} 
  \left[U^{(2)}_{n \Omega q}(r,z) \ U^{(1)*}_{n \Omega q}(r,z)
  + L^{(2)}_{n \Omega q}(r,z) \ L^{(1)*}_{n \Omega q}(r,z) \right] \ .
\end{eqnarray}
Similarly, starting from definitions (\ref{eq:kinetic_energy}) and (\ref{eq:current_divergence}), 
we obtain expressions for the kinetic energy density and the divergence of the current 
\begin{eqnarray}
 \tau_q(r,z) = \frac{1}{2 \pi}
 \left(2 \sum_{\Omega>0}^{\Omega_{max}} \right) \sum_{E_n>0}^{E_{max}}
 \left[  
   \frac{(\Omega - 1/2)^2}{r^2} \left| U^{(2)}_{ n \Omega q} \right|^2
  +\frac{(\Omega + 1/2)^2}{r^2} \left| L^{(2)}_{ n \Omega q} \right|^2 \right. \nonumber\\  
\left.  +\left| \frac{\partial U^{(2)}_{ n \Omega q}}{\partial r}  \right|^2
  +\left| \frac{\partial L^{(2)}_{ n \Omega q}}{\partial r}  \right|^2
  +\left| \frac{\partial U^{(2)}_{ n \Omega q}}{\partial z}  \right|^2
  +\left| \frac{\partial L^{(2)}_{ n \Omega q}}{\partial z}  \right|^2 
 \right]\;\; ,
\end{eqnarray}
\begin{eqnarray}
\nabla \cdot {\bf J}_q( {\bf r} )  = 
\frac{1}{2 \pi}
\left(2 \sum_{\Omega>0}^{\Omega_{max}} \right) \sum_{E_n>0}^{E_{max}}
  2  \left[
\frac{\partial U^{(2)}_{n \Omega q}}{\partial r} 
\frac{\partial L^{(2)}_{n \Omega q}}{\partial z}
- \frac{\partial L^{(2)}_{n \Omega q}}{\partial r} 
\frac{\partial U^{(2)}_{n \Omega q}}{\partial z}  \right.
&+&  \frac{\Omega - 1/2}{r} U^{(2)}_{n \Omega q} 
\left( \frac{\partial U^{(2)}_{n \Omega q}}{\partial r} -
 \frac{\partial L^{(2)}_{n \Omega q}}{\partial z}  \right) \nonumber\\
&-& \left. \frac{\Omega + 1/2}{r} L^{(2)}_{n \Omega q} 
\left( \frac{\partial U^{(2)}_{n \Omega q}}{\partial z} +
 \frac{\partial L^{(2)}_{n \Omega q}}{\partial r}  \right)  
\right].
\end{eqnarray}
\end{widetext}
The total number of protons or neutrons is obtained by integrating their
densities
\begin{equation}
N_q = \int d^3r \ \rho_q({\bf r}) = \left(2 \sum_{\Omega>0}^{\Omega_{max}} \right)
      \sum_{E_n>0}^{E_{max}} N_{n \Omega q} 
\end{equation}
with
\begin{equation}
N_{n \Omega q} = \int_0^{\infty} r dr \int_{-\infty}^{\infty} dz
\left[|U^{(2)}_{n \Omega q}(r,z)|^2 + |L^{(2)}_{n \Omega q}(r,z)|^2 \right]
\label{eq:dens_norm}      
\end{equation}
which gives the contribution of the quasiparticle state $| n \Omega q >$ to the proton
or neutron density. In the HF+BCS limit, $N_{n \Omega q} \rightarrow v^2_{n \Omega q}$.
An analogous treatment of the pairing density yields
\begin{equation}
P_q = \int d^3r \ \tilde{\rho}_q({\bf r}) = \left(2 \sum_{\Omega>0}^{\Omega_{max}} \right)
              \sum_{E_n>0}^{E_{max}} P_{n \Omega q}
\end{equation}
with the ``pairing density spectral distribution'' (with respect to energy and angular
momentum)
\begin{eqnarray}
P_{n \Omega q} = - \int_0^{\infty} r dr \int_{-\infty}^{\infty} dz
  \left[ U^{(2)}_{n \Omega q}(r,z) \ U^{(1)*}_{n \Omega q}(r,z) \right. \nonumber \\
\left. \ + \ L^{(2)}_{n \Omega q}(r,z) \ L^{(1)*}_{n \Omega q}(r,z)
   \right] \ .
\end{eqnarray}
In the HF+BCS limit, $P_{n \Omega q} \rightarrow (u \cdot v)_{n \Omega q}$.
Finally, we state the normalization condition for the four-spinor quasiparticle wavefunctions
as
\begin{equation}
\int d^3r  \ \psi^\dagger_{n \Omega q}({\bf r}) \ \psi_{n \Omega q}({\bf r}) \ = \ 1 \ ,
\end{equation}
which leads to 
\begin{eqnarray}
\int_0^{\infty} r dr \int_{-\infty}^{\infty} dz
\left[|U^{(1)}_{n \Omega q}(r,z)|^2 + |L^{(1)}_{n \Omega q}(r,z)|^2 \right. \nonumber \\
\left. \ + \ |U^{(2)}_{n \Omega q}(r,z)|^2 + |L^{(2)}_{n \Omega q}(r,z)|^2 \right]        
\ = \ 1 \ .
\end{eqnarray}

\section{\label{sec:numerical}Lattice representation of spinor wavefunctions and Hamiltonian}

For axially symmetric nuclei, we diagonalize the HFB Hamiltonian (\ref{eq:hfb2d})
separately for fixed isospin projection $q$ and angular momentum
quantum number $\Omega$.  We solve the eigenvalue problem by direct diagonalization on a 
two-dimensional grid $(r_\alpha,z_\beta)$, where $\alpha = 1,...,N_r$
and $\beta = 1,...,N_z$. (Because the grid usually extends from $-z$ to $+z$, we have
$N_r$ $\approx$ $N_z$, so when referring to the number of points in the mesh we only 
mention the values of $N_r$).
The four components of the spinor wavefunction 
$\psi(r,z)$ are represented on the two-dimensional lattice by an expansion in 
Basis-Spline functions $B_i (x)$ evaluated at the lattice support points. 
Further details about the Basis-Spline technique are given 
in Refs. \cite{Uma91,WO95}. For the lattice representation of the Hamiltonian,
we use a hybrid method \cite{K96,KO96,Obe99} in which
derivative operators are constructed using the Galerkin method; this 
amounts to a global error reduction. Local potentials are
represented by the Basis-Spline collocation method (local error reduction).
The lattice representation transforms the differential operator
equation into a matrix form
\begin{equation}
\sum_{\nu=1}^N {\cal{H}}_{n}^{\ \nu} \psi^{\Omega}_{\nu} = 
            E^{\Omega}_{n} \psi^{\Omega}_{n} \ \ \ (n=1,...,N) \ ,
\end{equation}
The HFB calculations are initialized using the density output from a prior
{\it HF+BCS} run which results in fast convergence of the HFB code.
Because the HFB problem is self-consistent we use an iterative method for the
solution, and at every iteration the full HFB Hamiltonian is diagonalized. 
Typically 15-20 iterations are sufficient for convergence at the level of one part
in $10^5$ for the total binding energy.
The Fermi levels $\lambda_q$ for protons and neutrons are calculated in every iteration by means of a
simple root search using the equations \cite{DF84}
\begin{eqnarray}
f(\lambda_q) &=& \bar{N}_q(\lambda_q) -N_q = 0 \nonumber \\
\bar{N}_q(\lambda_q) &=& \left(2 \sum_{\Omega>0}^{\Omega_{max}} \right)
      \sum_{E_n>0}^{E_{max}} \bar{N}_{n \Omega q}(\lambda_q) \nonumber \\
\bar{N}_{n \Omega q}(\lambda_q) \ &=& \ \frac{1}{2}
   \left[ 1-\frac{({\cal E}_{n \Omega q}-\lambda_q)}{(({\cal E}
_{n \Omega q}-\lambda_q )^{2}+\Delta^2_{n \Omega q})^{\frac{1}{2}}}\right] \nonumber \\
\Delta_{n \Omega q} \ &=& \ 2\ E_{n \Omega q}\sqrt{N_{n \Omega q}(1-N_{n \Omega q})}  \ ,
\label{eq:number_eq}
\end{eqnarray}
where $E$ denotes the quasiparticle energy, and ${\cal E}$ is the equivalent
single-particle energy (as defined by the BCS formalism). The quantity $N$ in the
last line of the equation denotes the spectral norm of the density as defined in
Eq. \ref{eq:dens_norm}.
The calculated value for $\lambda_q$ is used in the next iteration. This process is 
repeated until convergence is achieved.


\section{\label{sec:parameters} numerical parameters: $^{22} O$ calculations}

In this section, we present a series of studies of the numerical parameters in axially
symmetric HFB calculations. In particular, we study the dependence of observables on
the equivalent single particle energy cut-off, the lattice box size, the number of
mesh points, and the maximum angular momentum quantum number $\Omega_{max}$.
The numerical tests are carried out for $^{22}O$. This neutron-rich isotope has an
$N/Z$ ratio of $1.75$ and is close to the experimentally confirmed dripline nucleus
$^{24}O$. 

\begin{figure}[htb]
\vspace*{0.2cm}
\includegraphics[scale=0.37]{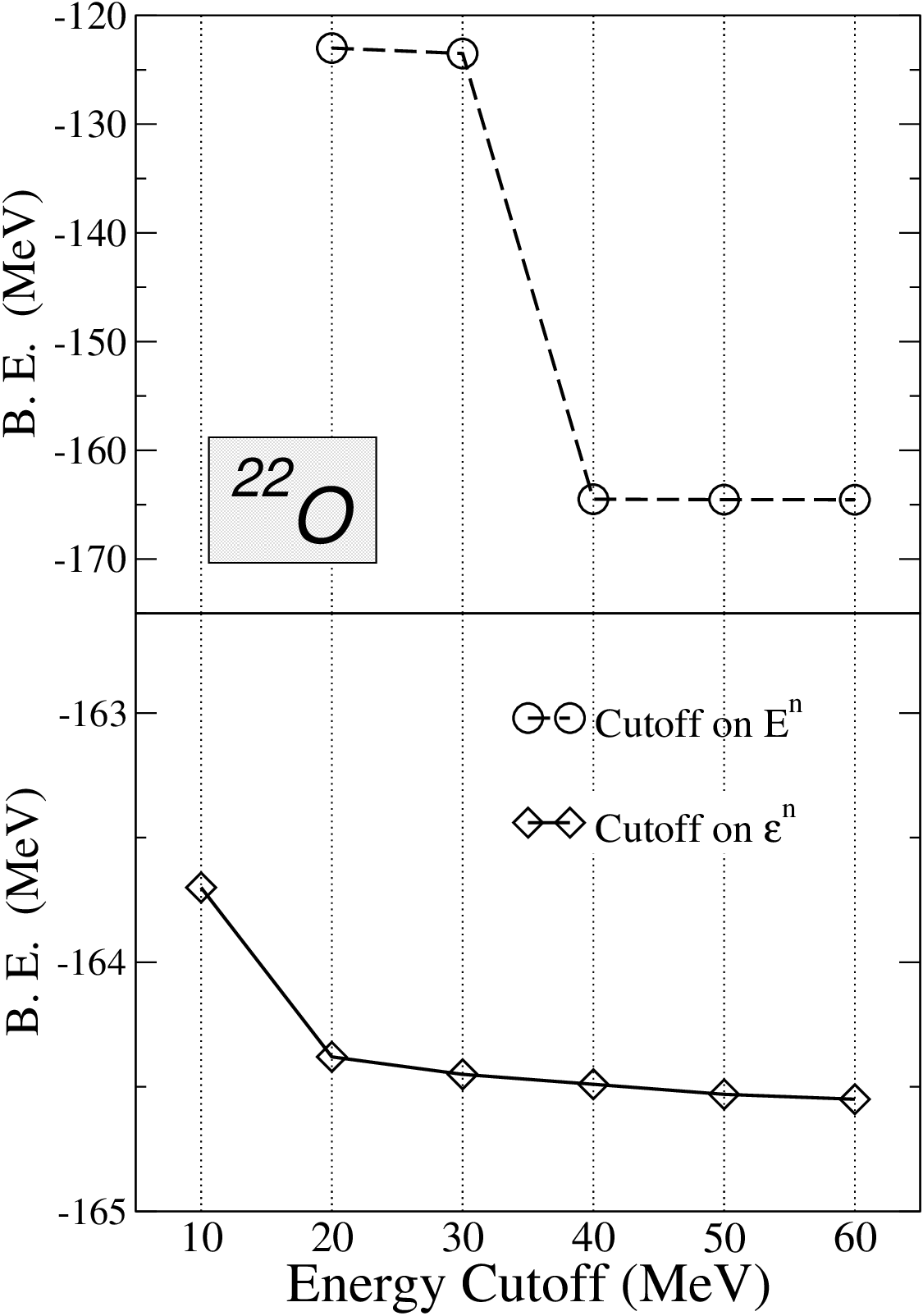}
\vspace*{0.0cm}
\caption{\label{fig:cutoff}  Binding energy of $^{22}O$ vs. energy cutoff. 
Top: cutoff in the quasiparticle spectrum, bottom: cutoff in the equivalent
single particle spectrum (absolute value). All calculations were performed 
with B-Spline order $M = 7$, $N_r = 18$ lattice points, angular momentum projection
$\Omega_{max} = \frac{5}{2}$ and box size $R = 10 fm$.}
\end{figure}

\subsection{Energy cutoff}

The numerical solution of the HFB equations on a 2-D lattice results in a set of 
quasiparticle wavefunctions and energies. The quasiparticle energy spectrum contains
both bound and (discretized) continuum states. The number of eigenstates is
determined by the dimensionality of the discrete HFB Hamiltonian,
which is $N = (4 \cdot N_r \cdot N_z)^2$, for fixed isospin projection $q$ and 
angular momentum projection $\Omega$. In our calculations, we typically
obtain quasiparticle energies up to about 1 GeV. It is well-known that zero-range
pairing forces require a limited configuration space in the $p-p$ channel
because the interaction matrix elements decrease too slowly with excitation energy
\cite{DN96}. One therefore introduces an energy cut-off, either in the 
quasiparticle energy ($E_{max}$) or in the equivalent single particle energy
(${\cal E}_{max}$). Hence, in the case of zero-range pairing forces the infinite
summations over quasiparticle energies in the expressions for the densities 
$\rho$, $\tau$, and current $J$ are terminated at a maximum quasiparticle energy. 

The quantity $E_{max}$ has to be chosen such that the maximum quasiparticle energy
exceeds the depth of the mean field nuclear potential, and all of 
the bound states have to be included in the sums \cite{DF84}.
We follow the prescription of Refs. \cite{DF84,SD00} to set the cutoff 
energy in terms of the equivalent single particle energy spectrum, ${\cal E}_{n}$. 
For the Skyrme SLy4 force with pure delta-pairing, Dobaczewski et al. \cite{DovPr} deduced
a pairing strength of $V_0=-170 MeV fm^3$, with ${\cal E}_{max}=60$ MeV. 
We utilize the same parameters in all of our 2-D calculations.

Even though ${\cal E}_{max}$ is a fixed parameter in the HFB calculations,
it is interesting to analyze the sensitivity of observables to the value
of the energy cutoff. In Fig.~\ref{fig:cutoff} we plot the total
nuclear binding energy for cutoff values of ${\cal E}_{max}$ between 10 and 60 MeV
and the same for ${E}_{max}$ from 20 to 60 MeV.
We find that in both cases, the binding energy remains essentially constant 
for cutoff values of 40 MeV and above. Clearly, a cut-off below 40 MeV results 
in significant changes in the binding energy because quasiparticle levels with 
large occupation probabilities are left out. This result is in agreement with 
the 1-D radial calculations of Ref.\cite{DN96}. 

\begin{figure}[htb]
\vspace*{0.3cm}
\includegraphics[scale=0.35]{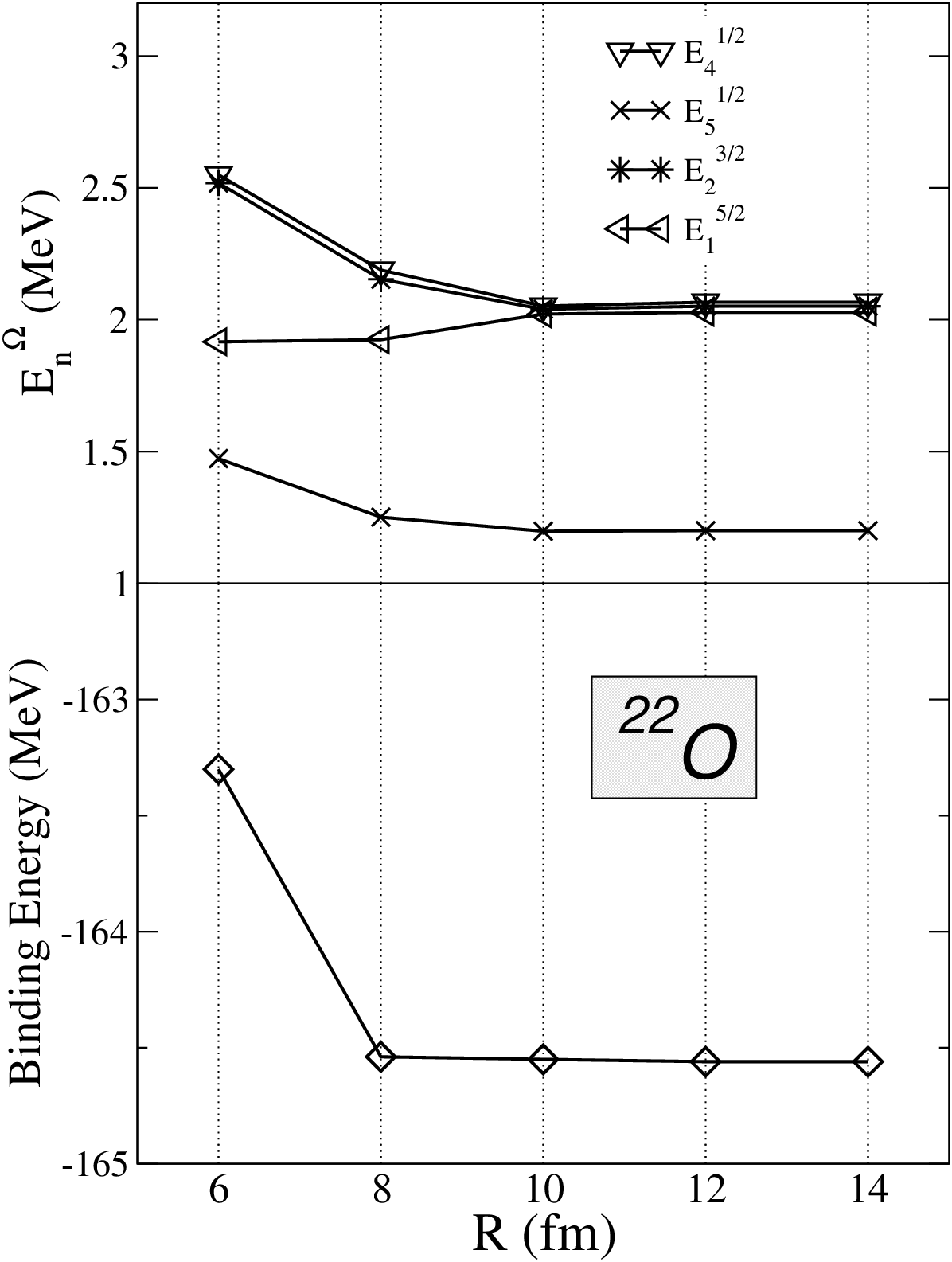}
\vspace*{0.0cm}
\caption{\label{fig:box} Bottom: Total binding energy of $^{22}O$ as a function
of the box size $R$. Top: Quasiparticle energies for states with large
occupation probability ($N_n$) as a function of $R$.
The spline order used was $M = 9$, $N_r = 19$ grid points, $\Omega_{max} = \frac{9}{2}$,
and cutoff energy ${\cal E}_{max}=60$ MeV.}
\end{figure}

\subsection{Lattice box size}
Using cylindrical coordinates, the lattice box size $R$ defines the boundary in
radial ($r$) direction; the box size in $z$ direction is $2R$. The value of $R$
must be chosen large enough for the wavefunctions to vanish at the outer edges
of the box and needs to be adjusted for optimal accuracy and computing time.
Figure \ref{fig:box} shows the dependence of the binding energy on $R$ for $^{22}O$. 
The mesh spacing was kept at a constant value of $\Delta r \approx 1 fm$. 
Figure \ref{fig:box} also presents some of the quasiparticle energy levels
$E_n^\Omega$ with large occupation probability $N_n$; these levels correspond to
low-lying states in the equivalent single-particle spectrum. Evidently,
the quasiparticle energies and the total binding energy converge in
essentially the same way with increasing box size. Figure \ref{fig:box}
shows that convergence is reached at {R}=10 fm. The behavior of the quasiparticle
states with respect to the mesh boundaries has also been discussed in Ref. \cite{DN96}.
For heavier systems, the box size has to be increased. A safe initial guess for $R$
is about three times the classical nuclear radius. Tests also show that
one may utilize the same mesh spacing for both light and heavy nuclei.

\begin{figure}[htb]
\vspace*{0.5cm}
\includegraphics[scale=0.40]{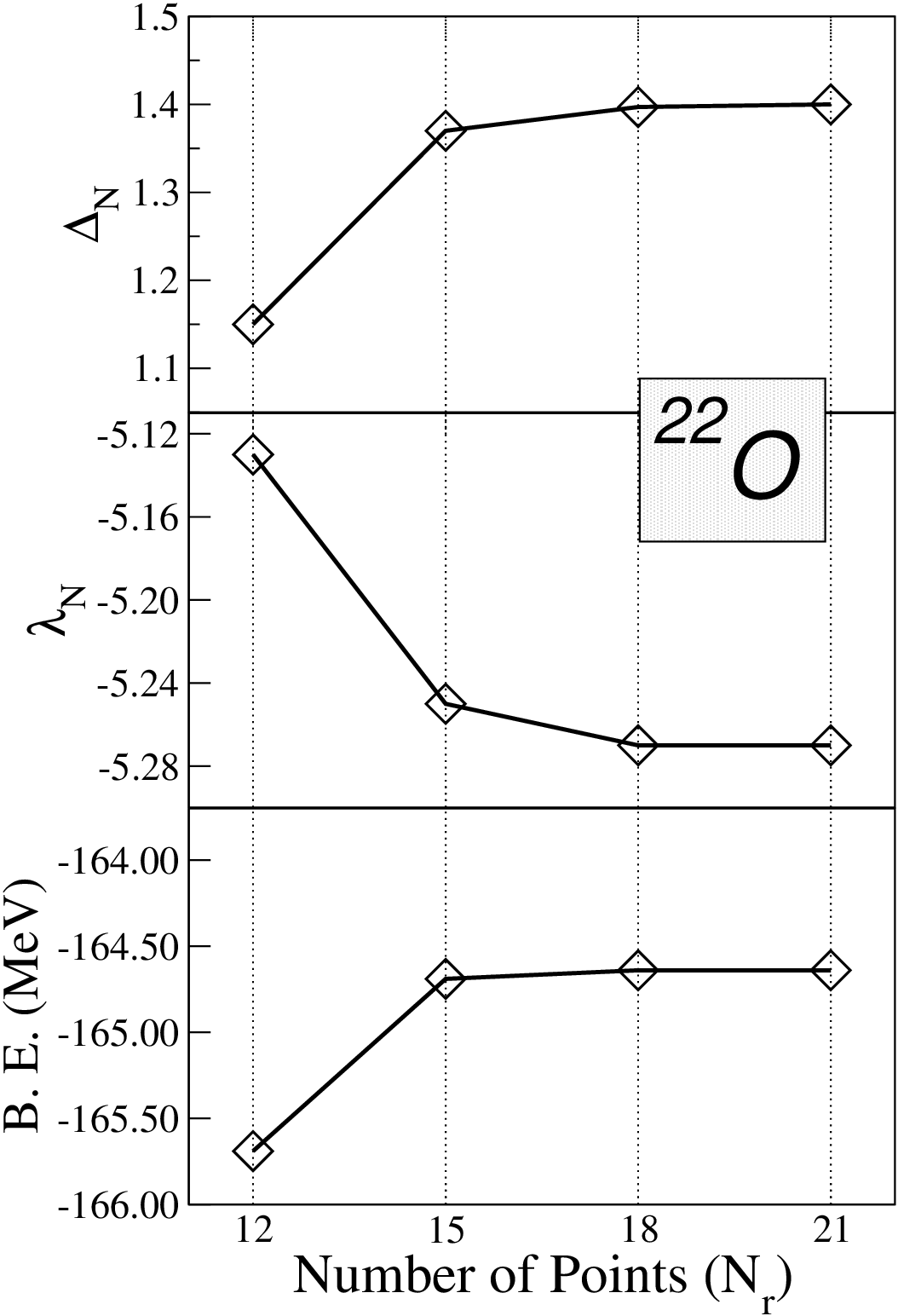}
\vspace*{0.0cm}
\caption{\label{fig:points} Total binding energy, Fermi level and pairing gap
for neutrons in $^{22}O$ vs. number of mesh points in radial direction, for
fixed box size $R = 8$ fm. The quantities $\Omega_{max}$ and ${\cal E}_{max}$
are the same as in Fig. 2}
\end{figure}

\subsection{Number of mesh points}
One of the major advantages of the B-Spline technique is that one can utilize a
relatively coarse grid resulting in a lattice Hamiltonian matrix of low dimension.
Fig. \ref{fig:points} shows several observables as a function of the number of radial mesh
points, for a fixed box size $R=8$ fm. The binding energy, neutron Fermi level, and pairing
gap for $^{22}O$ reach their asymptotic values at about 18 grid points in radial direction. 
For the fixed $(r,z)$ boundary conditions utilized in our work, the B-Spline lattice
points show a (slightly) non-linear distribution, with more points in the vicinity of 
the boundaries. In the central region, the grid spacing for 18 radial points is $0.75$ fm. 

\subsection{Projection of the angular momentum, $\Omega$}
\begin{figure}[htb]
\vspace*{0.3cm}
\includegraphics[scale=0.40]{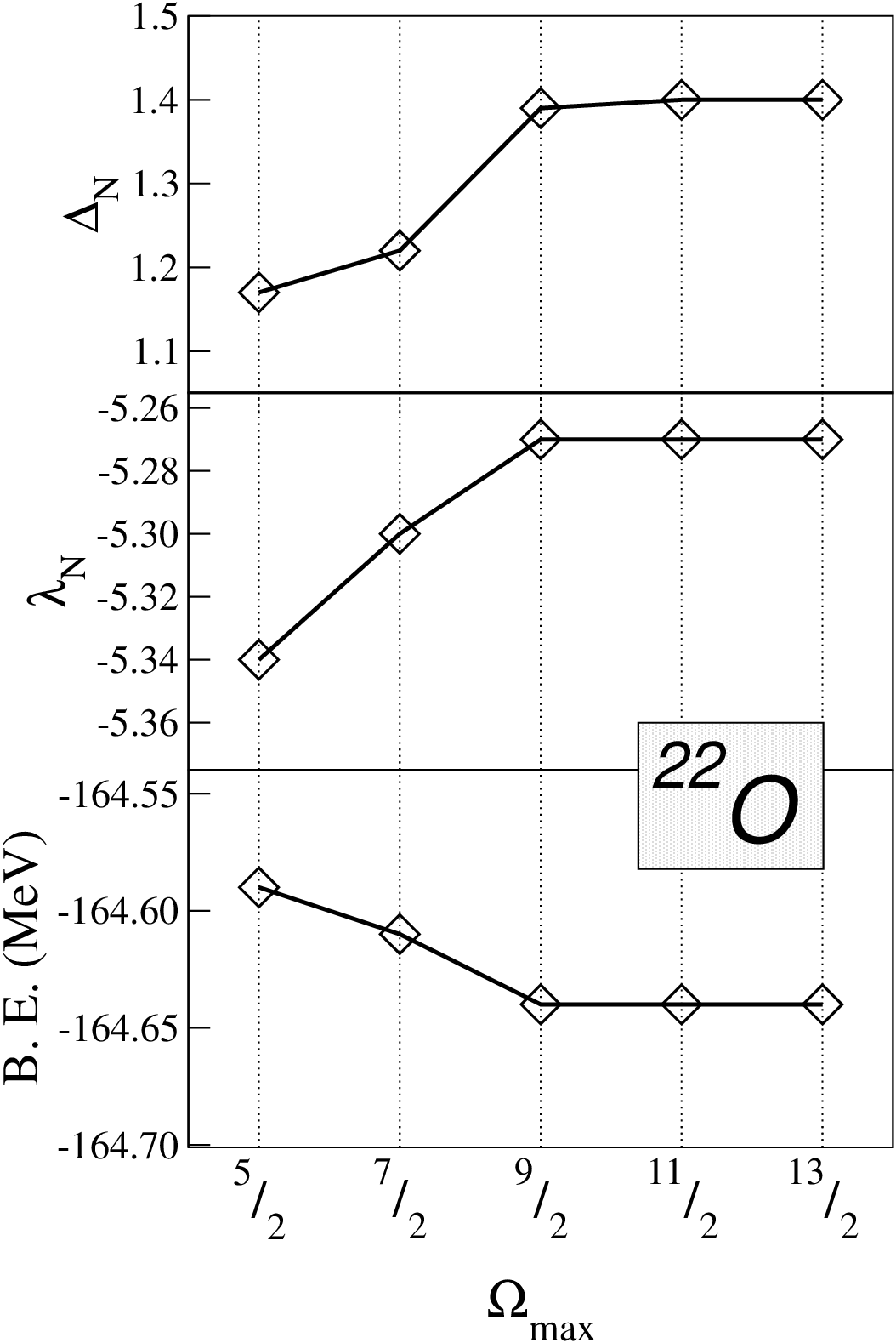}
\vspace*{0.0cm}
\caption{\label{fig:jzs_three} Binding energy, neutron Fermi level, and 
average neutron pairing gap for $^{22}O$ vs. maximum angular momentum
projection $\Omega_{max}$. Box size $R = 10$ fm, $N_r = 18$ and
an energy cutoff of 60 MeV were used.}
\end{figure}

It has been mentioned in the formalism section that all observables can be 
expressed by sums over {\it positive} $j_z$ quantum numbers $\Omega > 0$.
The maximum value $\Omega_{max}$ increases, in general, with the number of protons
and neutrons ($Z,N$) and also depends on the nuclear deformation. There is no
\textit{a priori} criterion to fix $\Omega_{max}$; this numerical parameter
needs to be determined from test calculations in various mass regions.
We have performed calculations for $^{22}O$ using $\Omega_{max}$ values from 5/2 to 13/2. 
Figure \ref{fig:jzs_three} displays the results for the total binding energy,
neutron Fermi energy and neutron pairing gap. All three observables converge
at $\Omega = 9/2$. 

\section{Results}
In this section we present converged numerical results of our 2D-HFB code.
Our main goal is to demonstrate the accuracy of our Basis-Spline expansion technique
on a 2-D coordinate lattice by comparison with the 1-D coordinate space results
of Dobaczewski et al. \cite{DN96,DovPr} for spherical nuclei. For this purpose we have
chosen two very neutron-rich spherical nuclei: a light nucleus, $^{22}O$, with
$N/Z=1.75$ and a heavy system, $^{150}Sn$, with $N/Z=2.0$. Finally, we will also
present results for a strongly deformed medium-heavy system, $^{102}Zr$
with $N/Z=1.55$. This system was chosen because it allows us to compare our lattice results
(which treat the continuum states accurately) to the ``transformed harmonic
oscillator'' (THO) expansion technique recently developed by Stoitsov et al.
\cite{SD00}. In this framework, a local-scaling point transformation of the spherical
harmonic oscillator is used to expand the quasiparticle wavefunctions in a 
set of bound single-particle wavefunctions.  

\begin{table}[b!] 
\caption{\label{table:comparison_O22}
Calculations for $^{22} O$ for HFB+SLy4.
The axially symmetric calculations (2D) of this work used a box size 
$R = 10 fm$ with maximum $\Omega = \frac{9}{2}$ 
and an energy cutoff of 60 MeV.
The spherical calculation of Ref. \cite{DovPr} was made with $R = 25 fm$ and a 
$j = \frac{21}{2}$. All calculations were made with a cutoff at 60 MeV.}
\begin{ruledtabular}
\begin{tabular}{ l c c c}
               &   1-D \cite{DovPr} & 2-D (THO)\cite{StoPr}  & 2-D (this work)  \\
\hline 
B. E.   (MeV)       & -164.60 & -164.52 & -164.64  \\
$\lambda_n$ (MeV)   &  -5.26  & -5.27   & -5.27  \\ 
$\lambda_p$ (MeV)   &  -18.88 & -18.85  &  -18.16 \\ 
$\Delta_n$ (MeV)    & 1.42    & 1.41    &  1.40 \\ 
$\Delta_p$ (MeV)    & 0.00    & 0.00    &  0.00 \\ 
$R_{rms}$  (fm)     &  2.92   & 2.92    &  2.92 \\
$\beta_2$           &   *     & 0.00002 &  0.0008 \\
$E_{pair}(n)$ (MeV) &  -2.85  & -2.78   & -2.75 \\
\end{tabular}
\end{ruledtabular}
\end{table}

\subsection{Exotic spherical nuclei: $^{22}O$ and $^{150}Sn$}
In Table \ref{table:comparison_O22} we compare our 2-D HFB results
for the spherical isotope $^{22}O$ with the 1-D radial HFB method of Ref.\cite{DF84}.
Corresponding results in the 2-D THO basis with 20 oscillator shells are also given.
All calculations were performed with the Skyrme SLy4 force in the p-h
channel and a pure delta interaction (strength $V_0= -170 MeV fm^3$) in the
p-p channel, corresponding to volume pairing. The table lists several observables:
the total binding energy (for comparison, the experimental value
is $-162.03$MeV), the Fermi level for protons and neutrons, 
the neutron energy gap (for protons, the gap is exactly zero in all three
calculations), the $rms$ radius, and the quadrupole deformation (note 
that both 2-D calculations predict essentially zero deformation).
Overall, the results of the axially symmetric code of the present work agree 
with the other two calculations in all the observables. The binding
energy predicted by our 2D-lattice code is very close (within 40 keV) to
the 1-D lattice result, while the THO method result differs by 80 keV.
The difference in the Fermi level for protons is due to different
conventions in choosing this energy for magic numbers. We choose the
Fermi energy to be the mid point of the energy of the last occupied
level and the first unoccupied level.

\begin{table}[t!]
\caption{\label{table:comparison_Sn150}
Comparison of calculations for spherical nucleus $^{150} Sn$ with HFB+SLy4.
The 1-D calculations were made by Ref. \cite{DovPr}, using a 
box size $R = 30$ and a linear spacing of points of 0.25 fm, with $j_{max}$ of 
$\frac{21}{2}$.
Calculations by the axially symmetric HFB 2-D code were made using a box size 
$R = 20 fm$ with $N_r = 23$, maximum $\Omega = \frac{13}{2}$. In both calculations 
the pairing strength $V_0$ was set to -170 $MeV fm^3$, and the energy 
cutoff to 60 $MeV$.}
\begin{ruledtabular}
\begin{tabular}{ccc}
 &1-D&2-D\\ 
\hline 
B. E  (MeV)      & -1129.0  &  -1129.6  \\
$\lambda_n$ (MeV)& -0.96  &  -0.94  \\ 
$\lambda_p$ (MeV)& -17.54 &  -17.69 \\ 
$\Delta_n$  (MeV)&  1.02  &   1.00  \\ 
$\Delta_p$ (MeV) &  0.00  &   0.02  \\
$R_{rms}$  (fm)  &  5.12  &  5.13   \\ 
$\beta_2$        &   *    &  0.01  \\ 
$E_{pair}(n)$ (MeV) & -10.452 & -10.057 \\
\end{tabular}
\end{ruledtabular}
\end{table}

We now present results for the tin isotope $^{150}Sn$, a heavy nucleus
far away from the valley of $\beta$-stability which is close to the two-neutron
drip-line. Table \ref{table:comparison_Sn150} gives a comparison
of our 2-D results (which predict a very small quadrupole deformation
$\beta_2=0.005$) with Dobaczewski's 1-D radial HFB calculations.
The box size used in the axially symmetric calculations was 20 fm in $r$ 
direction and 40 fm in the $z$ axis, whereas the 1-D code had a 30 fm radial box. 
Also, the density of points has a different meaning in the radial code, since 
it uses a different grid than the one used in the B-Splines technique for our 2-D 
code. For these calculations the resulting mesh spacing
in the 1-D code was 0.25 fm, whereas the maximum mesh spacing in the 2-D one 
was 1.1 fm. In the 2-D calculations an approximately $3000\times 3000$ matrix 
was diagonalized for each $\Omega$ and isospin value, and for each major
HFB iteration. The full calculation required about 30 HFB iterations.
Like in the oxygen isotope, the agreement is very good; a possible source of
small discrepancies is the fact that our 2-D code yields $\beta_2=0.005$
whereas the 1-D code \emph{assumes} an exactly spherical shape.
Table \ref{table:comparison_Sn150} also contains another interesting
piece of information on $^{150} Sn$: the neutron Fermi level $\lambda_n$
is located less than 1 MeV below the continuum which shows the proximity of
this nucleus to the two-neutron dripline.

\begin{table}[!b]
\caption{\label{table:zr_comparison}
Comparison of calculations $HFB+SLy4$ for $^{102} Zr$ with two different 
methods in the axial symmetry.
The configurational space calculations (THO) were made by Ref. \cite{StoPr}
with 20 oscillator shells and pairing strength of -187.10 $MeV fm^3$.
Calculations by the coordinate space HFB 2-D code were made using a box size 
$R = 12 fm$ with $N_r=19$, maximum $\Omega = \frac{11}{2}$, $V_0$  -170 
$MeV fm^3$ and the energy cutoff of 60 $MeV$.}
\begin{ruledtabular}
\begin{tabular}{ l c c}
                  & 2-D (THO)  & 2-D (this work)  \\
\hline 
B. E.   (MeV)     & -859.40    & -859.61  \\
$\lambda_n$ (MeV) & -5.42      & -5.46    \\ 
$\lambda_p$ (MeV) & -12.10     &  -12.08  \\ 
$\Delta_n$ (MeV)  & 0.56       &  0.31    \\ 
$\Delta_p$ (MeV)  & 0.62       &  0.34    \\ 
$R_{rms}$  (fm)   & 4.58       &  4.58    \\ 
$\beta_2$         & 0.429      &  0.431   \\ 
\end{tabular}
\end{ruledtabular}
\end{table}

\subsection{\label{sec:tin} Deformed neutron-rich nucleus: $^{102} Zr$}
Our main motivation for developing an axially symmetric code is to
perform highly accurate calculations for deformed nuclei, including the
continuum states. The zirconium isotope $^{102} Zr$ is a heavy nucleus with strong prolate
quadrupole deformation in its ground state. Its neutron to proton ratio of $N/Z=1.55$
places it into the neutron-rich domain although it is likely far away from the neutron
dripline (in the 1-D spherical HFB+SkP approximation \cite{SD93} the last bound nucleus
in the chain is predicted to be $^{136} Zr$). We have chosen this isotope primarily because
our results can be compared to the stretched harmonic oscillator expansion (THO) method
mentioned above which does not involve any continuum states.

In Table \ref{table:zr_comparison} we present the results of our 2-D HFB calculations
in coordinate space with the results obtained by the THO method. A comparison of the
total binding energy of the system in both methods shows a difference of about
21 keV which can be considered small in comparison to the absolute value of the energy
(the experimental binding energy value is $-863.7$ MeV). The pairing strength 
parameter, $V_0$, used in each calculation also makes a difference. Other observables
(Fermi levels, rms-radius and deformation $\beta_2$) agree quite well, also. 
However, we find differences in the energy gap values ($\Delta_n$, $\Delta_p$);
these may be attributed to the different density of states used in the two methods
see Eq. (\ref{eq:mean_delta}) or to the extrapolation of the oscillator
parameter in the THO approach to deformed systems.

\section{\label{sec:conclusions}Conclusions}

In this paper, we have solved for the first time the \emph{HFB continuum problem in coordinate
space for deformed nuclei in two spatial dimensions} without any approximations.
The novel feature of our new HFB code is that it takes into account high-energy
continuum states with an equivalent single-particle energy of 60 MeV or more.
In the past, this has only been possible in 1-D calculations for spherical nuclei
\cite{DN96}. Current 3-D HFB codes in coordinate space, e.g. Ref. \cite{TH96}, utilize
an expansion of the quasiparticle wavefunctions in a truncated HF-basis 
which is limited to continuum states up to about $5$ MeV of excitation energy.
 
The Vanderbilt HFB code has been specifically designed to study ground state
properties of deformed axially symmetric even-even nuclei near the neutron
and proton drip lines. 
The large pairing correlations near the driplines and the strong coupling to
the continuum represent major challenges for the numerical solution.
We have solved the HFB problem on a two-dimensional grid in cylindrical
coordinates $(r,z)$ using a Basis-Spline representation of wavefunctions and operators.
B-Splines are a generalization of the well-known finite element technique.
By using B-Splines of order $M=9$ (corresponding to polynomials of up to 8-th order)
we are able to represent derivative operators very accurately on a relatively
coarse grid with a lattice spacing of about 0.8 fm. 
While our current 2-D lattices are linear, a major advantage of the B-Spline
technique is that it can be extended to nonlinear lattices \cite{K96,KO96} which will be
particularly useful for an accurate and efficient calculation of neutron
skins in heavy nuclei.

In this work, we have used the Skyrme (SLy4) effective N-N interaction in the p-h channel, 
and a pure delta interaction (corresponding to volume pairing) in the p-p channel. We present results for 
binding energies, deformations, normal densities and pairing densities, Fermi levels,
and pairing gaps.

We have investigated the numerical convergence of several observables as a function
of lattice box size, grid spacing, angular momentum $\Omega_{max}$, and we
have studied the sensitivity of the observables to the continuum cut-off. These test calculations
were carried out for the neutron-rich isotope $^{22}O$ with $N/Z=1.75$ which is 
close to the dripline nucleus $^{24}O$. 

Our HFB-2D code predicts a spherical shape for the neutron-rich nuclei $^{22}O$
and $^{150}Sn$. In this case, our calculations can be compared with the 
1-D radial HFB results of Dobaczewski et al. \cite{DN96}, and indeed there
is good agreement between the two. We also present results for a strongly
deformed system, $^{102} Zr$, in which case we present a comparison with
the stretched oscillator expansion method of Stoitsov et al. \cite{SD00,StoPr}.

We have implemented our code on an IBM-SP massively parallel supercomputer.
Parallelization is possible for different angular
momentum states $\Omega$ and isospins (p/n). 
We will also study alternative numerical techniques, in particular damping methods
which we have utilized for solving the Dirac equation on a 3-D lattice \cite{WO95}.

In the near future, we plan to investigate several isotope chains, with particular
concentration on deformed nuclei. We also plan to study
a variety of Skyrme parameterizations for the mean field, and both
volume and surface pairing.
As more data from existing RIB facilities become available, it is likely that 
it will become necessary to develop new effective N-N
interactions to describe these exotic nuclei. Furthermore, our 2-D HFB 
groundstate wavefunctions can be used as input into coordinate-space based QRPA
calculations \cite{Ma01} to investigate collective excited states of nuclei
near the driplines.

\begin{acknowledgments}
This work is supported by the U.S. Department of Energy under grant No.
DE-FG02-96ER40963 with Vanderbilt University. Some of the numerical calculations
were carried out on CRAY and IBM-SP supercomputers at the National Energy Research
Scientific Computing Center (NERSC). We also like to acknowledge many
fruitful discussions with W. Nazarewicz and M. Stoitsov (ORNL) and with
J. Dobaczewski (Warsaw).
\end{acknowledgments}

\begingroup
\squeezetable
\begin{table*}
\caption{\label{tab:new_skyrme_parameters} Skyrme force parameters. Values for 
new parameters $b_0$, $b'_0$, $b_1$, $b'_1$, $b_2$, $b'_2$, $b_3$, $b'_3$, $b_4$, 
and $b'_4$ have been calculated using relations \ref{parameters_transformation} and
\ref{extra_transformation}, from old parameterization \cite{VB72},\cite{BQBGH82},
\cite{RD99}. Numbers have been rounded to three decimal places.}
\begin{ruledtabular}
\begin{tabular}{ldddddddddd} 
Force  & b_0 & b'_0 & b_1 & b'_1 & b_2  & b'_2 & 
        b_3 & b'_3 & b_4 & b'_4 \\ \hline
	
SkM*        & -2764.025   & -1560.55   & 68.75  & 68.125   & 170.625  & 68.437  
              & 3898.75     &1949.375   &65.0    &65.0  \\   
$Z_{\sigma}$  &-3145.945  &-3316.251 &64.495  &58.315    &148.877 &61.405  
              &5577.823   &6707.621  &61.845  &61.845 \\
SkT6        &-2145.863  &-1600.426 &0.0     &0.0       &110.25   &0.0    
              &4005.312   &3204.25    &53.5    &53.5 \\
SLy4        &-3526.790  &-3320.210 &32.484 &-49.289  &185.325  &62.665 
              &5776.007   &6385.639  &61.5    &61.5 \\
SkI1        &1000.310   &869.809   &32.354 &-49.803  &-432.059 &-1136.719  
              &580.693    &-2810.714 &62.13   &62.13 \\
SkI3        &-2034.628  &-1424.936 &32.301 &-127.914  &100.074  &-124.799 
              &3336.309   &3632.793  &94.254  &0.0  \\
SkI4        &-2231.708  &-1679.676 &32.271 &-75.310  &-121.462  &-528.369 
              &3814.977   &3991.101  &183.097 &-180.351  \\
SkP         &-3359.948  &-2322.346 &44.642 &89.284   &190.343   &140.223 
              &5100.600   &3185.341  &50.0    &50.0  \\
SkO         &-1882.032  &-608.585  &22.537 &15.075   &-72.754   &-358.023 
              &2660.027   &237.585   &176.578 &-198.749  \\
SkO'        &-2068.449  &-987.770  &19.156 &8.312    &41.250   &-128.648 
              &3132.384   &1192.344  &143.895 &-82.889  \\ 
\end{tabular}
\end{ruledtabular} 
\end{table*}
\endgroup


\appendix*
\section{\label{sec:skyrme}Skyrme parameterization}
The (density dependent) two-body effective N-N interaction is given by
\begin{eqnarray*}
v^{(2)}_{12} &=& t_{0} \ (1+x_{0}\Ph_{\sigma }) \ \delta ({\bf r}_{1}-{\bf r}_{2}) \nonumber\\
&+& \frac{1}{2} \ t_{1} \ (1+x_{1}\Ph_{\sigma }) \ \{ \delta ({\bf r}_{1}-{\bf r}_{2})\kh^{2} ) + 
\kh'^{2}\delta ({\bf r}_{1}-{\bf r}_{2})\} \nonumber\\ 
&+& t_{2} \ (1+x_{2}\Ph)\ \vchk' \cdot \delta ({\bf r}_{1}-{\bf r}_{2})\ \vchk \nonumber\\
&+& \frac{1}{6} \ t_{3} \ (1+x_{3}\Ph_{\sigma })\ \rho ^{\gamma }\ \delta ({\bf r}_{1}-{\bf r}_{2}) \nonumber\\ 
&+& i\ W_{0} \ (\sh_{1}+\sh_{2})\cdot \{\vchk'\times\delta ({\bf r}_{1}-{\bf r}_{2})\vchk \} \ ,
\end{eqnarray*}
\noindent $\Ph$ being the exchange operator, and $\vchk,\vchk'$ relative momentum operators. This form of the 
interaction with parameters $x_{0}, x_{1},x_{2}, x_{3},t_{0}, t_{1}, t_{2},t_{3}, t_{4}$, has being changed to an equivalent one with 
$b_1,b_1', b_2,b_2', b_3,b_3', b_4,b_4'$, parameters \cite{RD99}. This is done through the transformation
\begin{equation}
\left( 
\begin{array}{c}
t_1\\ 
t_1 x_1 \\
t_2\\
t_2 x_2
\end{array}
\right) =
\left( 
\begin{array}{rrrr}
4/3&8/3&-2/3&-4/3\\ 
-2/3&-4/3 &4/3&8/3\\ 
4&-8/3&2&-4/3\\ 
-2&4/3&-4&8/3 
\end{array}
\right) 
\left( 
\begin{array}{c}
b_1\\ 
b_2\\
b_1'\\
b_2'
\end{array}
\right) \ .
\label{parameters_transformation}
\end{equation}
and 
\begin{eqnarray}
t_0 &=& \frac{4}{3}b_0-\frac{2}{3}b_0' \nonumber \\
t_0 x_0 &=& -\frac{2}{3}b_0+\frac{4}{3}b_0' \nonumber \\
t_3 &=& \frac{16}{3}b_3-\frac{8}{3}b_3' \nonumber \\
t_3 x_3 &=& -\frac{8}{3}b_3 + \frac{16}{3}b_3' \nonumber \\
t_4 &=& 2 b_4 =  2 b_4' \;\;.
\label{extra_transformation}
\end{eqnarray}
The last equation only holds for certain forces, as shown in Table 
\ref{tab:new_skyrme_parameters}. For forces like SKI and SKO $b_4$ and $b_4'$
get different values.


\subsection{Energy density}
Calculation of the energy expectation value for an arbitrary interaction involves
carrying out an integration over six dimensions in coordinate space.  One of the primary 
advantages of an interaction that contains a delta function, like the Skyrme one, is that 
the evaluation of such integral becomes substantially simplified, and it's reduced to a 
three-dimensional evaluation 
\begin{equation}
\label{energy_density_integral}
E = \langle \Phi | H | \Phi \rangle = \int d^3 r \; {\cH}(\vcr) \ .
\end{equation}
The Hamiltonian density {\cH}(\vcr) is composed of several terms
\begin{equation}
\cH = \cH_0 + \cH_{LS} + \cH_C \;\;.
\end{equation}
The kinetic energy and some of the density dependent 
terms in the Skyrme interaction are included in
\begin{eqnarray}
\lefteqn{
\cH_0 = \frac{\hbar^2}{2m}\ \tau + \frac{b_0}{2} \ \rho^2
- \frac{b'_0}{2}  \ \sum_q \rho_q^2 } \nonumber \\
\;\;\;\;\;&+& \frac{b_3}{3} \ \rho^{\alpha + 2}
- \frac{b'_3}{3} \ \rho^{\alpha}\sum_q \rho_q^2  \nonumber \\
&+& b_1 \left(\rho \tau - j^2 \right)
- b'_1  \sum_q \left( \rho_q \tau_q - j_q^2 \right) \nonumber \\
&-& \ \frac{b_2}{2} \ \rho \nabla^2 \rho \ + 
\frac{b'_2}{2} \  \sum_q \rho_q \nabla^2 \rho_q
\;\;.
\label{eq:h_0}
\end{eqnarray}
The current densities ( $\vcj$, $\vcj_q$) appearing in this term are 
identically zero for time independent states. 
The finite range spin-orbit terms have the form
\begin{equation}
\cH_{LS} = - \ b_4 \ \rho {\bf \nabla \cdot J} - b'_4 \ \sum_q \rho_q ({\bf \nabla \cdot J}_q) \;\;.
\end{equation}
The Coulomb term contains an integral over the proton density as well as the Slater exchange
term,
\begin{eqnarray}
\cH_C = 
\frac{e^2}{2} \int d^3 r' \rho_p(\vcr) \frac{1}{|\vcr - \vcr'|} \rho_p(\vcr') \nonumber\\
 - \frac{3}{4} e^2 \left( \frac{3}{\pi} \right)^{1/3}
\left[ \rho_p(\vcr) \right]^{4/3}\;\;.
\end{eqnarray}


\subsection{Single Particle Hamiltonian}
The Hartree--Fock Hamiltonian using the Skyrme effective interaction can be written as
\begin{eqnarray}
\label{eq:hf_hamiltonian}
h_q = &-& {\bf\nabla  \cdot} \frac{\hbar^2}{2 m_q^*} {\bf \nabla} 
    + U_q + U_C \cdot \delta_{q,p} \nonumber\\
    &+& \frac{1}{2 i} \left( {\bf \nabla \cdot I}_q 
   + {\bf I}_q{\bf \cdot \nabla}
    \right)  - i {\bf B}_q{\bf \cdot}\left( {\bf \nabla \times \sigma} \right)\;\;.
\end{eqnarray}
Several effective quantities appear in this equation. The effective mass is defined by   
\begin{equation}
\label{effective_mass_def}
\frac{\hbar^2}{2 m_q^*} = \frac{\hbar^2}{2m} 
 \ + \ b_1 \ \rho  \ - \ b'_1 \ \rho_q  \;\;,
\end{equation}
the effective current density
\begin{equation}
\label{effective_current_def}
{\bf I}_q = - \ 2 \ b_1 \ {\bf j} \ + 2 \ b'_1 \ {\bf j}_q\;\;,
\end{equation}
and the effective spin density
\begin{equation}
\label{spin_orbit_potential}
{\bf B}_q = b'_1 \ {\bf J}_q \ + \ b_4 \ \nabla \rho \ + \  b'_4 \ \nabla \rho_q 
\;\;.
\end{equation} 
As previously indicated, all of the terms in Eq.(\ref{effective_current_def})
vanish for bound states.  Also, the first term in Eq.(\ref{spin_orbit_potential})
is usually ignored. \\

The effective nuclear potential for the Skyrme force is given by 
\begin{eqnarray}
U_q &=& b_0 \  \rho \ - \ b'_0 \ \rho_q +  b_1 \ \tau - b'_1 \ \tau_q \nonumber \\
      &+& \frac{b_3}{3}  \ \left( \alpha  + 2 \right) \rho^{\alpha + 1} 
      - \frac{b'_3}{3}  \ \left[ \alpha \rho^{\alpha - 1} \sum_q \rho_q^2 
          + 2 \rho^{\alpha} \rho_q  \right] \nonumber \\
      &-& b_4 \ \nabla \cdot {\bf J} - b'_4 \ \nabla \cdot {\bf J}_q  \nonumber \\ 
      &+& b'_2 \ \nabla^2 \rho_q \ - \ b_2 \ \nabla^2 \rho\;\;,
\end{eqnarray}
and the Coulomb field is
\begin{equation}
U_C = e^2 \int d^3 r' \frac{\rho_p({\bf r'})}{|{\bf r} - {\bf r'}|} 
    - e^2 \left(\frac{3}{\pi} \right)^{1/3}\left[\rho_p({\bf r})\right]^{1/3}\;\;.
\end{equation}
${\cal B}_r$ and ${\cal B}_z$ from equations (\ref{eq:spin_orbit_operator}) for 
the spin-orbit part representation of the potential operator are given by:
\begin{subequations}
\begin{eqnarray}
{\cal B}_r \equiv {\bf B_q \cdot e_r} 
&=& \nabla_r (b_4 \rho + b'_4 \rho_q) \\
{\cal B}_z \equiv {\bf B_q \cdot e_z}
&=& \nabla_z (b_4 \rho + b'_4 \rho_q) \ ,
\end{eqnarray}
\end{subequations}
$b_4$ and $b'_4$ values are shown in Table \ref{tab:new_skyrme_parameters} 
for different forces.


\end{document}